# Graphene Based Quantum Dots

H. G. Zhang[1,*], H. Hu[1,*], Y. Pan[1,*], J. H. Mao[1], M. Gao[1], H. M. Guo[1], S. X. Du[1], T. Greber[2,†], and H.-J. Gao[1,†]

[1] Institute of Physics, Chinese Academy of Sciences, P.O. Box 603, Beijing 100190, China

[2] Physik-Institut, University of Zürich, Winterthurerstrasse 190, CH-8057 Zürich, Switzerland

**Abstract:**

**Laterally localized electronic states are identified on a single layer of graphene on ruthenium. The individual states are separated by 3 nm and comprise regions of about 90 carbon atoms. This constitutes a quantum dot array, evidenced by quantum well resonances that are modulated by the corrugation of the graphene layer. The quantum well resonances are strongest on the isolated "hill" regions where the graphene is decoupled from the surface. This peculiar nanostructure is expected to become important for single electron physics where it bridges zero-dimensional molecule-like and two-dimensional graphene on a highly regular lattice.**

* Those authors contribute equally to this work.

† Corresponding authors: greber@physik.uzh.ch, hjgao@iphy.ac.cn.

Graphene is *the* emerging material for nanoscale devices. As a single sheet of carbon, graphene, has unique chemical and physical properties. It is a honeycomb network where strong in-plane $sp^2$ hybrid bonds guarantee chemical stability and the interference of the remaining $p_z$ electrons in the honeycomb is responsible for physical properties like that of a gapless semiconductor with a point-like Fermi surface. This implies a low conduction electron density with a large electron mobility *(1)* allowing the construction of new transistors *(2)* and in combination with magnetic fields the observation of e.g. the quantum Hall effect at room temperature *(3)*. If the honeycomb symmetry is broken, the physical properties change. This is the case for graphene on a substrate, a situation that occurs in most production processes, like e.g. the segregation from SiC single crystals *(4)*, or the chemical vapor deposition schemes, where hydrocarbon gases as e.g. ethylene ($C_2H_4$) on transition metals *(5-9)*. Similar to hexagonal boron nitride layers, where the symmetry in the honeycomb is intrinsically broken due to the differences between boron and nitrogen, the substrate breaks the symmetry of the graphene honeycomb ($C_A$, $C_B$) and the formation of gaps is observed, where the choice of the substrate allows the control of the magnitude of the gap, and its position relative to the Fermi level *(10)*. Also, in plane boundaries of the $sp^2$ hybridized layers impose a distortion of the properties of the free standing layers, which is known from one-dimensional graphene nanoribbons *(11)*. For zero dimensional graphene, i.e. small graphene flakes or polycyclic aromatic hydrocarbons *(12)*, where benzene is the smallest representant, non-dispersing, localized electronic states with a gap that decreases with the number of carbon rings are expected. However, not much is known about the transition from a localized, dot like hexagonal carbon network to a delocalized one, also because it is difficult to prepare "zero dimensional" graphene.

Here we focus on the observation and identification of quantum dots on a single layer of graphene. They emerge as a regular array on a 3 nm superlattice with about 90 carbon atoms or a diameter of 2 nm. This peculiar structure is formed in a self-assembly process, when graphene is grown on a Ru(0001) surface *(6-9)*. The lattice mismatch of about 10% between the substrate and the graphene leads to the formation of a super structure, where N+1 graphene lattice constants fit on N substrate lattice constants. For g/Ru(0001) N is 11.5 and a corresponding moiré pattern is expected *(14)*. The anisotropic bonding of the C atoms with respect to the underlying Ru atoms leads, however, to a strong lateral lock in energy, where C on top of Ru is preferred. This causes dislocations and a deviation from a regular moiré structure: The carbon layer and the substrate are strained anisotropically and the much softer out of plane elastic modulus of graphene causes a strong corrugation of the carbon sheet. Regions where none of the two carbon atoms in the honeycomb bond on top of a Ru atom form a "mound-" or "hill-"like protrusion (see Fig. 1A). The height of the protrusion is about 0.1 nm as measured with scanning tunneling microscopy *(6-9)* or found from density functional theory calculations *(15)*, and affects the properties of the structure dramatically. A height histogram of the carbon atoms in the supercell indicates that 30%, i.e. about 90 carbon atoms have a height larger than 0.05 nm, which we assign to the hill. It has been shown that the hills have a 0.25 eV higher local work function *(16)* compared to the "valley" regions, where the graphene sheet is closely bound to the substrate. A splitting in the C 1s core level photoemission spectrum confirmed two species of carbon atoms on g/Ru(0001) that were related to the corrugation, where about one third has a 0.6 eV lower binding energy *(17)*. On the other hand, in the valence bands no such splitting could be found. Only one dispersing π band with a relatively large gap was observed *(16)*. This seeming paradox of absence of a corrugation induced splitting in the valence band may be resolved if we assign to the hills a molecule like behavior as isolated quantum dots, without dispersion. The isolation of these dots and the concomitant electronic states are related to the corrugation of the structure, where the lift off of the hills causes vertical and lateral localization. The vertical localization arises from the interface and is pronounced by the decoupling of the graphene from the substrate, while the lateral localization is ascribed to a confinement due to the corrugation of the graphene. The quantum dots show up as sharp resonances in the conductance spectra in low temperature scanning tunneling microscopy experiments.

Fig. 1 shows high resolution scanning tunneling microscopy and spectroscopy data from g/Ru(0001) recorded at 4 K. In Fig.1A the large-scale image of the super structure with a lattice constant of 3 nm is shown. The single layer graphene spans across two terraces separated by an atomic step. The inset is a high-resolution zoom in, where the dotted line shows the cut on which individual scanning tunneling spectra are measured. The *dI/dV* conductance spectra recorded with an average tunneling current of 100 pA are shown as an $U$ vs. $x$ map in Fig.1B. The color code represents the conductance from the tip into unoccupied states of g/Ru(0001). Clearly, a series of resonances at distinct tunneling voltages is observed. The energies and the sharpness of the resonances change within the 5 nm cut across the super-cell. One of these peaks, the second lowest one, shows a behavior that deviates from the others, which are the well known field emission resonances, ubiquitous at tip-surface junctions *(18, 19)*. The field emission resonance energies may be used to determine the local work function of surfaces, whereby a decrease in energy indicates a decrease of the work function of the probed surface region *(20, 21)*. For the present case g/Ru(0001) the energy up-shift on the hill confirms the local work function shift as found by photoemission from adsorbed xenon *(16)*. The peak that opposes the trend of the field emission resonances is, as it is shown here, a quantum well resonance, which indicates the quantum dot nature of the hills. It can be seen that this resonance undergoes within less than 1 nm an abrupt decrease in energy (0.5 eV) in going from the valley to the hill of the super structure, which conveys an isolated nanoscopic electronic system on the hills that act like "mesas". The conductance map is not fully symmetric with respect to the top of the hills. This indicates that the valleys on the left and to the right are distinct, as it is reflected in the coordination of the $C_A$ and $C_B$ atoms to *fcc* and *hcp* sites respectively *(6, 16)*. Fig. 1C shows two spectra, one on a hill and one in a valley, at positions $x$=0 and $x$=2.0 nm, respectively. The field emission resonances are labeled with *FER n*. The label *QWR* is used for quantum well resonance. The quantum well resonance on the hills displays a sharp peak and evidences a high Q-factor of the resonance.

In the following we want to substantiate the physical picture leading to the quantum dots reflected in the site dependent spectra in Fig. 1B, and explain the opposite trend between the field emission resonances and the quantum well resonance. Fig. 2 shows a one-dimensional model of the potential between the surface and the STM tip for the tip/g/Ru(0001) system with positive bias voltages, where electrons tunnel from the tip to the sample and the concomitant solutions of the one-dimensional Schrödinger equation. In Fig. 2A the potential in the valley and in Fig. 2B that on the hills are depicted. In the valley the local work function and the graphene-ruthenium distance is lower than that on the hill.

The model is a potential well system where the solutions of the Schrödinger equation are eigenstates with energies that correspond to the observed resonances. The fact that the experimentally observed peaks have a line width is not explained by the experimental resolution only, but indicates a finite lifetime that is related to the scattering of the electrons into other states, and will be further discussed below. The potential in the vacuum is modeled as the work function plus an image potential $U(z)$ prop. $1/4(z-z_i)$, with $z_i$ being the position of the image potential plane, plus a linear term that mimics the potential gradient between the tip and the sample. For the applied measurement mode this gradient is not exactly proportional to the applied tunneling voltage since the tunneling current is kept constant and thus the tip-sample distance increases for increasing tunneling voltage. The essential ingredient of the model, the interface between the graphene and the vacuum, is described with a delta-function potential centered on the image plane $\gamma\delta(z-z_i)$. This barrier accounts for the reflectivity of low energy electrons approaching the surface along $z$ *(8)*. The graphene is modeled as a rectangular quantum well with a width $a$ and $a'$, and a depth $V_o$ and $V_0'$ for hill and valley, respectively. The potential of the sample between $z_i$ and $z_i+z_m$ is set constant as the depth $V_o$. The local work functions $\Phi$ and $\Phi'$ are taken from the experiment *(16)*. The substrate is modeled as a perfect hard wall mirror, which is justified with the large gap in the relevant energy window along Gamma bar *(22)*. This potential-model is solved numerically by the Numerov algorithm *(23)*, and the parameters from the fit are shown in Table 1. The model reproduces the observed experimental trends of the up-shift of the field emission resonances and the down-shift of the quantum well resonance in going from the valley to the hill. The smaller potential $V_o'$ on the hills compared to $V_o$ in the valleys is taken as an indication of a different bonding of the graphene to the Ru. Of course, the well potential of the real system is more complex than that of the one-dimensional model. In particular the model does not account for the lateral localization. For an electron in a cylinder with a radius of 1.0 nm we expect an additional localization energy of 0.2 eV, which is not considered in our model, though this would mainly shift the value of $V_o'$.

The model also delivers the wave functions and thus allows the distinction of quantum well resonances and the field emission resonances by their locations. In Fig. 2C and D the amplitudes of the wave functions are plotted along *z*. The field emission resonances are mainly localized in the vacuum/graphene interface and the quantum well resonance is localized in the graphene/ruthenium quantum well. These states are the solutions of the 1D Schrödinger problem, with energies $E_n$ and wave functions $\Psi_n$, where *n* denotes the number of the nodes in (0, +∞). Without the graphene quantum well potential, the eigenstates $E_n$ are all field emission resonances (*FERs*) that show up as standing waves in the vacuum between tip and surface. When the graphene quantum well potential is added, the spectra alter depending on the intrinsic properties of the quantum well (i.e., *a*, $V_0$, *a'* and $V_0'$). Since one of the potential walls is penetrable, the quantum well state will hybridize with *FERs* with similar energy. This hybridization also implies that the *FERs* close to the *QWR* has amplitude in the well, and vice versa, the *QWR* gains amplitude outside the well, which is expected to have an effect on the *dI/dV* spectra. In Fig. 1C it can be seen that the second and the third peak on the blue curve collected at the valley region have a similar intensity and appear "quenched" with respect to the other peaks. The distinction between *FER* and *QWR* is not sharp, though it singles out the peculiar role of the *QWR* and the identification of quantum dots on g/Ru(0001). Also, a model without tip still predicts the quantum well resonance and the image potential states.

The quantum well resonance is reminiscent to the situation in a Fabry-Pérot interferometer *(24)*, or a resonance cavity, where the delta function of the vacuum/graphene interface acts as a semi-transparent mirror and the Ru substrate as a perfect mirror. Although the above physical picture also reminds to the so called transmission resonances *(25)*, we prefer the term quantum well resonance, since the states we observe have the nature of an electron in a Fabry-Pérot type interferometer where the electrons collect the phase in a multiple scattering process *(26)* in bouncing between the vacuum and the ruthenium, and not in a single scattering process as encountered in a transmission resonance *(27)*. This is further justified by the sharpness of the resonances. The increased conductance at the resonance energies thus does not result from direct transmission into the Ru substrate but from the increased density of states and the scattering of the electrons, from where they may join the conduction bands of the sample.

In Fig. 3 the experimentally observed resonances are compared to the model calculations. The intermediate energies between the valley and hill situation are interpolated linearly to the height as measured with the scanning tunneling microscope (Fig. 3B). The agreement is excellent and confirms the validity of the model.

Fig. 4A addresses the magnitude/width of the first field emission resonance (*FER 1*) and the quantum well resonance (*QWR*) on the 5 nm cut across the g/Ru(0001) super-cell. This further substantiates the quantum dot picture on the hills of g/Ru(0001). The quantum well resonance is strongest on the hills. Apparently the decoupling of the graphene from the substrate increases the quantum well resonance lifetime. In contrast, the field emission resonance *FER 1* on the hills is much weaker compared to the valleys. This is likely to have an electrostatic explanation, indicating the importance of the lateral dimensions: On the hills the convex shape in the electrostatic potential due to local work function differences defocuses the electrons out of the resonator cavity, while they are focused in the resonator cavity in the concave potential of the valleys. This explanation of the resonance width is illustrated in Fig. 4B where the influence of the topography and the concomitant electrostatic landscape is sketched for the tip/g/Ru(0001) tunneling junction.

In summary, the formation of graphene based quantum dots on Ru(0001) made from about 90 carbon atoms is demonstrated by means of scanning tunneling spectroscopy data with nanometer resolution. The observations are explained with a quantum well resonance and field emission resonances. The quantum well resonance on the hills of the corrugated graphene is very strong and has a distinctly lower energy compared to that in the valleys. It constitutes graphene based quantum dots, i.e. laterally and vertically confined states. This peculiar nanostructure is expected to become useful for nanoscience. The quantum dots are small enough that they are candidates for single electron physics at room temperature, where e.g. correlations between periodically arranged quantum dots shall be studied, or when molecules shall be self-assembled and isolated on a substrate. Also, in an applied magnetic field the spin splitting of the quantum well resonance should be observed. We expect that this structure is the first example only of periodic graphene based quantum dot systems and it has some potential for application in single electron quantum devices.

## Acknowledgements

Work at IOP was supported by grants from National Science Foundation of China, National "973" project of China, the Chinese Academy of Sciences, and the Sino-Swiss Science and Technology Cooperation IP09-092008.

**Figure Captions:**

FIG. 1. Scanning tunneling microscopy (STM) and spectroscopy of one layer of graphene on Ru(0001). (A) Large scale (120 nm × 120 nm) ($U_t$ = -2.0 V, $I_t$ = 0.1 nA) STM topographic image across two substrate terraces separated by a monoatomic step. The bright dots with a corrugation of 0.1 nm represent the hills in the superstructure with a lattice constant of ~3 nm. The inset shows the high-resolution image (6 nm × 6 nm), the dotted line indicates the cuts shown in (B). (B) Color scale map of the conductance ($dI/dV$) into the unoccupied substrate states as a function of tunneling voltage $U_t$ and position along the dashed line marked in the inset of (A). The center of the hill is taken as the zero position. (C) Conductance $dI/dV$ spectra on the hill $x$=0 (red) and the valley $x$=2.0 nm (blue). The field emission resonances are labeled as *FER n* ($n$=1, 2, 3, … 6). The quantum well resonance is labeled as *QWR*. The fingerprint of the quantum dot is the localized quantum well resonance at $x$~0±1 nm. Note the opposite energy shifts of the field emission resonances and the quantum well resonance. The spectra are taken at constant current $I_t$ = 0.1 nA.

FIG. 2. One-dimensional potential models for the quantum well resonance and the field emission resonances on g/Ru(0001). The essential ingredient is the delta potential $\gamma\delta(z-z_i)$ on the image plane at $z_i$ that accounts for the reflectivity of the electrons on the graphene surface. (A) For the valley region. (B) For the hill region. For the fitted values of the parameters see Table 1. The corresponding energies $E_n$ and amplitudes of the normalized wave functions $\Psi_n$ are indicated in (C) and (D). The quantum well resonances (red, *QWR*) are mainly localized in the interface between graphene and Ru and are sensitive to the interface.

FIG. 3. (A) Experimental (circles) and theoretical (triangles) resonance energies of the first three field emission resonances *FER n* ($n$= 1, 2, 3) and the quantum well resonance (*QWR*) as a function of the positions across the super-cell. The dashed lines are obtained by interpolating the theoretical results linearly to the line profile in (B). (B) The line profile shows the corrugation across the hill region of the graphene superstructure.

FIG. 4. (A) Maximum conductance of the first field emission resonance (*FER 1*) and the quantum well resonance (*QWR*) as function of the positions across the g/Ru(0001) super-cell. On the hills the conductance into the quantum well resonance is maximum, while the field emission resonance almost vanishes. (B) Schematic illustration of the tip-sample geometry that affects the intensities of the two differently located states in an opposite way.

Table 1. Parameters as obtained from the fit of the model in Fig. 2 to the observed energies of the field emission and the quantum well resonances in the valley and on the hill. For the valley $a + z_m$ was set close to the literature value of the graphene distance above the Ru substrate of 0.22 nm *(15)*, $V_0$, $z_0$ and $\gamma$ are left as free parameters. $z_o$ is a parameter that is needed in order to determine the potential gradients from the measured tunneling voltages and *relative* tip positions. It corresponds to the effective tip sample distance at a tunneling voltage of 2.0 V and with a current of 0.1 nA in the valley. On the hill $z_o'$ is set to the fit value from the valley and $V_0'$, and $\gamma'$ are left as free parameters. For the corrugation $a'-a$, a value of 50 pm has been assumed. The local work functions $\Phi$ and $\Phi'$ for both regions are taken from the experiment *(16)*.

# References and Notes

1. S.V. Morozov *et al*., *Phys. Rev. Lett*. **100**, 016602 (2008).

2. X. R. Wang *et al*., *Phys. Rev. Lett*. **100**, 206803 (2008).

3. K. S. Novoselov *et al*., *Science* **315**, 1379 (2007).

4. C. Berger *et al*., *Science* **312**, 1191 (2006).

5. A. T. N'Diaye *et al*., *Phys. Rev. Lett*. **97**, 215501 (2006).

6. S. Marchini, S. Günther, and J. Wintterlin, *Phys. Rev. B* **76**, 075429 (2007).

7. A. L. V´azquez de Parga *et al*., *Phys. Rev. Lett*. **100**, 056807 (2008).

8. P. W. Sutter, Jan-Ingo Flege *and* Eli A. Sutter, *Nature Mater*. **7**, 406 (2008).

9. Y. Pan *et al*., Adv. Mater. **21**, 2777 (2009).

10. G. Giovannetti *et al*., *Phys. Rev. Lett*. **101,** 026803 (2008).

11. X. Li et al., Science **319,** 1229 (2008).

12. M. Müller et al. Chem. Eur. J. **4,** 2099 (1998).

13. F. Schedin *et al*., *Nature Mater*. **6**, 652 (2007).

14. D. Martoccia *et al*., *Phys. Rev. Lett*. **101**,126102 (2008).

15 B. Wang *et al*., *Phys. Chem. Chem. Phys*. **10**, 3530 (2008).

16. T. Brugger *et al*., *Phys. Rev. B* **79**, 045407 (2009).

17. A. B. Preobrajenski *et al*., *Phys. Rev. B* **78**, 073401(2008).

18. R. S. Becker *et al*., *Phys. Rev. Lett*. **55**, 987 (1985).

19. G. Binnig *et al*., *Phys. Rev. Lett*. **55**, 991 (1985).

20. D. B. Dougherty *et al*., *Phys. Rev. B* **76**, 125428 (2007).

21. P. Ruffieux *et al*., *Phys. Rev. Lett*. **102**, 086807 (2009).

22. T. Pelzer *et al*., *J.Phys.:Condens. Matter* **12,** 2193 (2000).

23. D. Dunn *et al*., *J. Phys. A: Math. Gen.* **22**, L1093 (1989).

24. M. C. Tringides, M. Jalochowski and E. Bauer, *Phys.Today* **60**, 50 (2007).

25. J. A. Kubby, Y. R.Wang, and W. J.Greene, *Phys. Rev. Lett*. **65**, 2165 (1990).

26. N. V. Smith *et al*., *Phys. Rev. B* **49,** 332 (1994).

27. W. B. Su *et al*., *J. Phys.: Condens. Matter* **18,** 6299 (2006).

# Supplementary materials

**Experiments:**

Experiments were performed in an Omicron low temperature STM systems with a base pressure of the $10^{-10}$ mbar. The substrate ruthenium was cleaned by cycles of 1.0 keV Ar+ sputtering followed by annealing to 1350 K, and then oxidized at about 1000K with an $O_2$-pressure of $5.0\times10^{-7}$ mbar for about 3 minutes to remove the possible carbon impurities. Graphene layers on Ru (0001) were fabricated by exposing the substrate to ethylene at 1000K with the pressure of $5\times10^{-7}$ mbar for about 100 seconds. The sample was then cooled down to 4.5 K in the STM stage before scanning. Distance-voltage ($z$–$V$) spectroscopy were measured by positioning the tip over a fixed point on the surface and ramping the tip-sample bias with feedback engaged to maintain constant current. The tip retracts away from the sample in response to the voltage ramp was recorded as a function of voltage. The corresponding *dI/dV* spectra were acquired using a lock-in technique with a $10mV_{rms}$ sinusoidal modulation signal at 800 Hz. All the given bias voltage was referred to the sample with respect to the tip.

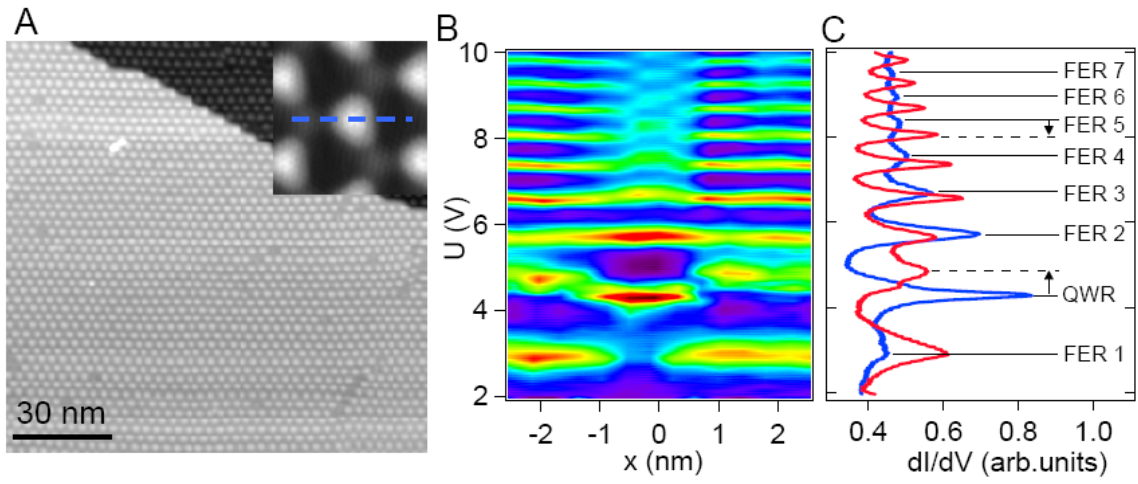

Figure 1

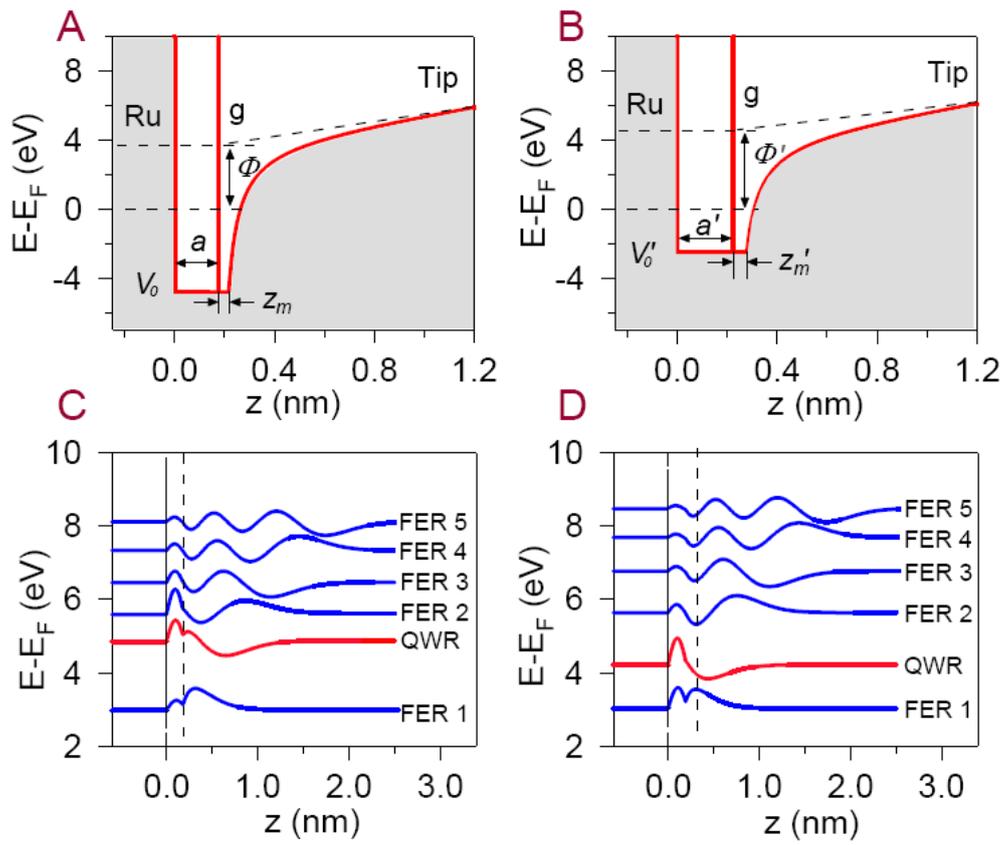

Figure 2

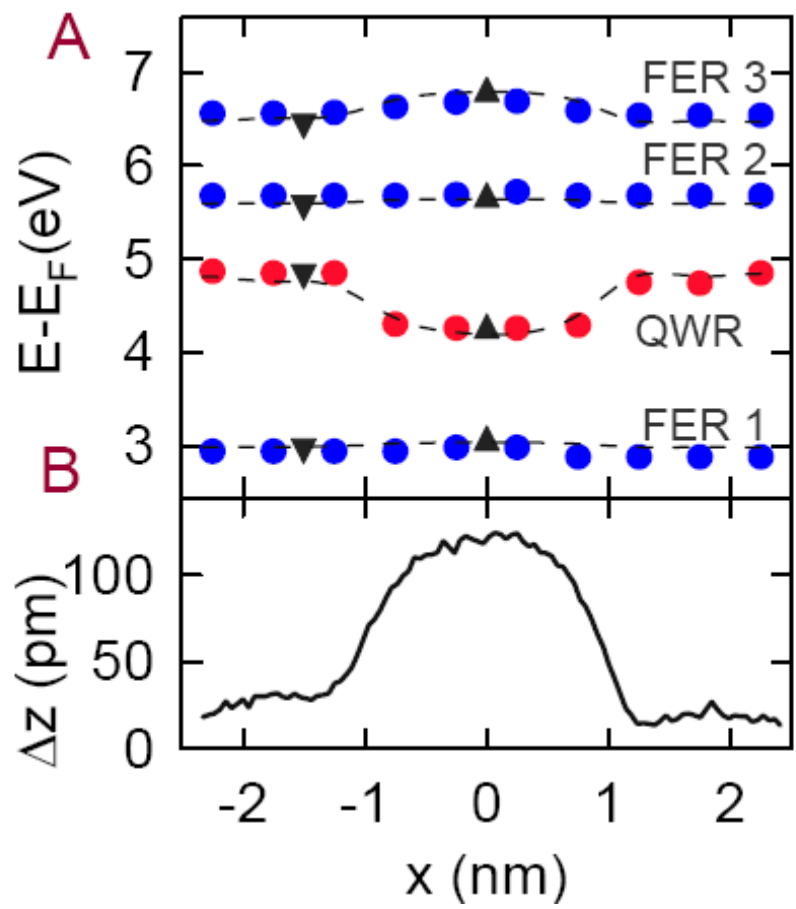

Figure 3

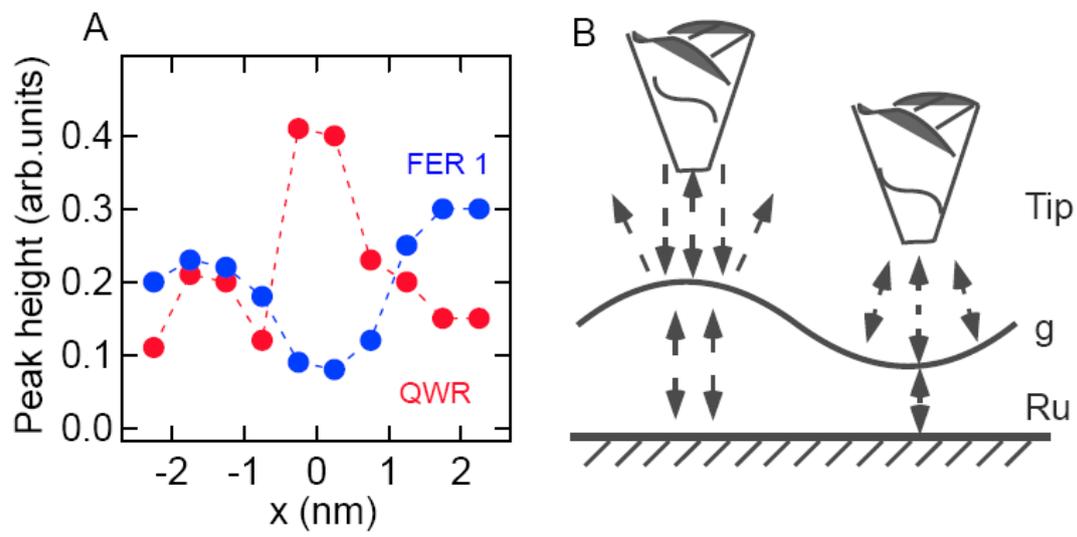

Figure 4

|  | $a$ (nm) | $V_0$(eV) | $\gamma$ (eV/Å) | $z_0$(nm) | $z_m$(nm) | $\Phi$ (eV) |
|---|---|---|---|---|---|---|
| Valley | 0.175 | -4.8 | 34 | 1.55 | 0.041 | 3.9 |
| Hill | 0.225 | -2.5 | 30 | 1.55 | 0.053 | 4.2 |

Table 1